\documentclass[aps,prl,showpacs,twocolumn]{revtex4}
\usepackage{graphicx} 
\usepackage{dcolumn} 
\usepackage{multirow}
\begin{document}

\title{Hexagonal Indium Double Layer on Si(111)-($\sqrt{7}\times\sqrt{3})$}

\author{Jae Whan Park}
\author{Myung Ho Kang}
\email{kang@postech.ac.kr}
\affiliation{Department of Physics, Pohang University of Science and Technology, Pohang 790-784, Korea}
\date{\today}

\begin{abstract}

Density-functional calculations are used to verify the atomic structure of the hexagonal In/Si(111)-($\sqrt{7}\times\sqrt{3}$) surface, which has been considered to represent an ultimate two-dimensional (2D) limit of metallic In overlayers.
Contrary to the prevailing assumption, this surface consists of not a single layer but a double layer of In atoms, which corresponds to a hexagonal deformation of the well-established $rectangular$ In double layer formed on Si(111)-($\sqrt{7}\times\sqrt{3}$) [Park $et$ $al$., Phys. Rev. Lett. 109, 166102 (2012)].
The same double-layer thickness accounts well for the typical coexistence of the hexagonal and rectangular phases and their similar 2D electronic structures.
It is thus conclusive that, regardless of rectangular or hexagonal, the In/Si(111)-($\sqrt{7}\times\sqrt{3}$) surface does not represent an one-atom-thick In overlayer.

\end{abstract}

\maketitle



One-atom-thick metal overlayers grown on semiconductor surfaces have attracted great attention as model systems ideal for exploring intriguing low-dimensional metallic properties \cite{carp96,yeom99,qin09,kim10,kwon14}.
One representative system of recent interest is the In/Si(111)-($\sqrt{7}\times\sqrt{3}$) surface, where the In overlayer was generally assumed to be one atom thick \cite{surn95,kraf95,kraf97} and so represent an ideal  two-dimensional (2D) limit of metallic In properties.
Fascinating electronic features of the In overlayer, including a nearly-free-electron Fermi surface,   \cite{rote03}, superconducting transitions \cite{zhan10,uchi11}, and an intriguing metallic transport behavior \cite{yama11}, have been explored and referred to as revealing the ultimate 2D limit. 
With little structural information about the In overlayer, however, its actual layer thickness has long been an open question.  


Particularly interesting in this regard is a double-layer picture for the In/Si(111)-($\sqrt{7}\times\sqrt{3}$) surface, contrary to the prevailing single-layer assumption.
In a recent density-functional theory (DFT) study \cite{park12}, of the two different In-derived ($\sqrt{7}\times\sqrt{3}$) phases distinguishable in scanning tunneling microscopy (STM) experiments \cite{surn95,kraf95,kraf97}, the more representative, rectangular phase (hereafter, $\sqrt{7}$-rect) was verified to contain an In double layer [see Fig. 1(a)]. 
Consequently, the remaining hexagonal phase ($\sqrt{7}$-hex) has been spotlighted as a single-layer alternative, because it typically developes a little earlier than the $\sqrt{7}$-rect phase when prepared by In deposition on the Si(111)-(7$\times$7) surface at about 400 $^{\circ}$C \cite{kraf95, kraf97}. 
Since the $\sqrt{7}$-hex phase evolves between the 4$\times$1 and $\sqrt{7}$-rect phases in the coverage-dependent growth process \cite{kraf97, yama13}, its In coverage would be in between 1.0 ML of the 4$\times$1 phase \cite{bunk99, cho01} and 2.4 ML of the $\sqrt{7}$-rect phase \cite{park12} (here, 1 ML refers to one In atom per surface Si), but the In coverage and thickness is yet to resolve.  


In early STM studies, Kraft $et$ $al.$ suggested a 1.0 ML single-layer model for the $\sqrt{7}$-hex phase \cite{kraf95,kraf97}, where the observed five STM protrusions per ($\sqrt{7}\times\sqrt{3}$) unit cell were attributed to five In atoms (corresponding to 1 ML).
This experimental model was more quantified by Shang $et$ $al.$ in a recent DFT study \cite{shan12}.
Another single-layer model with an In coverage of 1.2 ML was lately proposed by Uchida $et$ $al.$ \cite{uchi13a}, based on a similarity of their DFT simulations to the experimental STM image \cite{kraf95}.
The In coverage itself is also a matter of debate as seen in two recent experiments: while Uchihashi $et$ $al.$ \cite{uchi13} supported the 1.2 ML single-layer model of Uchida $et$ $al.$ \cite{uchi13a} based on the similarity of STM images, Yamada $et$ $al.$ \cite{yama13} estimated as about 2 ML in their reflection high-energy electron diffraction (RHEED) study.
Any single-layer model, however, has an inherent experimental problem: in their STM study, Kraft $et$ $al$. observed that the coexisting $\sqrt{7}$-hex and $\sqrt{7}$-rect phases have almost the same STM heights (with a small difference by 0.1--0.2 {\AA}) and also show very similar electron tunneling spectra \cite{kraf95}.
Previously, these similarities were argued to ensure that both phases have a single In layer \cite{kraf95}, but now the recent double-layer verification of the $\sqrt{7}$-rect phase \cite{park12} should imply a double-layer model for the $\sqrt{7}$-hex phase as well.


Here, we propose a new structural model for the hexagonal In/Si(111)-($\sqrt{7}\times\sqrt{3}$) surface.
This model features a double layer of 2.4 ML In atoms, which actually corresponds to a hexagonal deformation of the In double layer of the $\sqrt{7}$-rect phase. 
In what follows, the soundness of the hexagonal double-layer model is demonstrated by DFT calculations.

\begin{figure*}
\centering{ \includegraphics[width=16.0 cm]{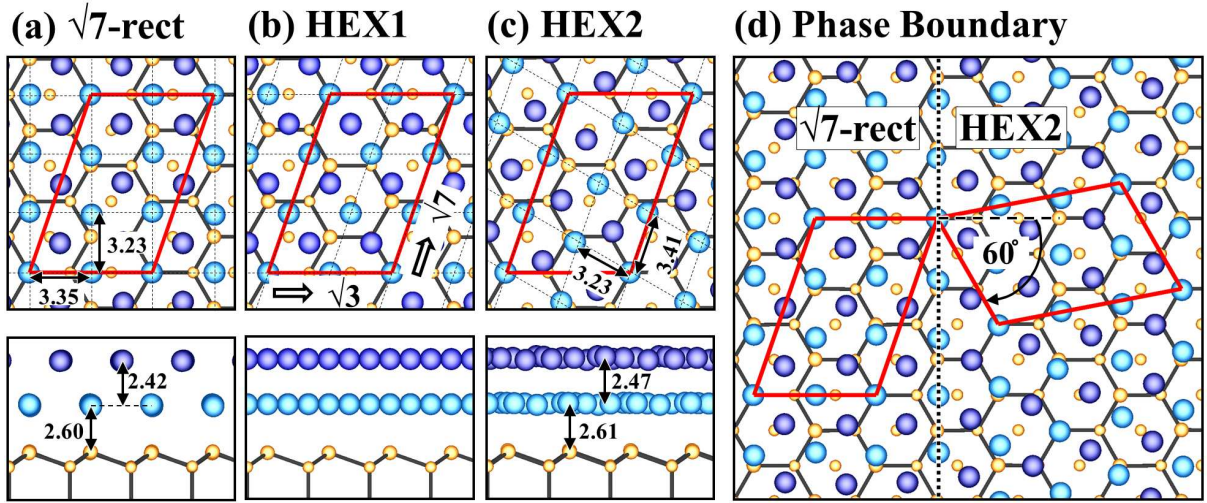} }
\caption{ \label{fig1}
(Color online)
Double-layer models for In/Si(111)-($\sqrt{7}\times\sqrt{3}$). 
(a-c) One rectangular (Ref. \cite{park12}) and two hexagonal models.  
Large (small) balls denote In (Si) atoms.
Solid and dashed lines denote ($\sqrt{7}\times\sqrt{3}$) and effective In-(1$\times$1) unit cells, respectively.
Numbers denote optimized bond lengths and layer spacings.
(d) Phase boundary between $\sqrt{7}$-rect and HEX2.
}
\end{figure*}


We perform DFT calculations using the Vienna $ab$-$initio$ simulation package within the generalized gradient approximation \cite{perd91} and the projector augmented wave method \cite{bloc94,Kres96}.
The Si(111) surface is modeled by a periodic slab geometry with six atomic layers and a vacuum spacing of about 12 {\AA}.
The calculated value 2.370 {\AA} is used as the bulk Si-Si bond length.
Indium atoms are adsorbed on the top of the slab, and the bottom is passivated by H atoms.
The electronic wave functions are expanded in a plane-wave basis with a kinetic energy cutoff of 246 eV.
A 4$\times$6$\times$1 $k$-point mesh is used for the ($\sqrt{7}\times\sqrt{3}$) Brillouin-zone integrations.
All atoms but the bottom two Si layers are relaxed until the residual force components are within 0.01 eV/{\AA}.


In light that the $\sqrt{7}$-hex and $\sqrt{7}$-rect phases have similar STM heights \cite{kraf95}, we examine first the possibility of hexagonal deformations of the double-layer $\sqrt{7}$-rect phase.
Figure 1 shows two easy-to-derive hexagonal models.
One is obtained from the $\sqrt{7}$-rect phase by a monoclinic deformation of the rectangular In array along the $\sqrt{3}$ direction, as captured in Fig. 1(b) (hereafter, HEX1).
The other (HEX2) is obtained from HEX1 by a monoclinic deformation along the $\sqrt{7}$ direction as shown in Fig. 1(c). 
In our calculations, while HEX2 is locally stable, HEX1 is unstable and relaxes back to the $\sqrt{7}$-rect phase with no energy barrier. 
We also examined variants of HEX1 and HEX2 by allowing lateral displacements of In atoms along the $\sqrt{3}$ or $\sqrt{7}$ direction, but all of them were found to converge to either the $\sqrt{7}$-rect phase or the HEX2 model.
Figure 1(c) shows the optimized HEX2 structure, which forms almost a regular In lattice ($a_{\rm 1}$ = 3.23 {\AA} and $a_{\rm 2}$ = 3.41 {\AA}): the In-In and In-Si interlayer spacings are 2.47 {\AA} and 2.61 {\AA} in average, respectively, comparing well with the values, 2.42 {\AA} and 2.60 {\AA}, of the $\sqrt{7}$-rect model. 
In energetics, the HEX2 model is as stable as the $\sqrt{7}$-rect model with a slightly higher formation energy by 0.05 eV per ($\sqrt{7}\times\sqrt{3}$) cell, being a promising model for the $\sqrt{7}$-hex phase. 

\begin{figure}
\centering{ \includegraphics[width=8.0 cm]{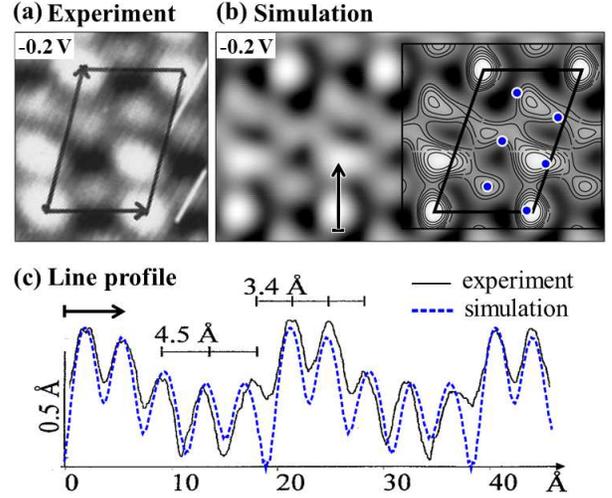} }
\caption{ \label{fig2}
(Color online)
STM comparison for the HEX2 model.  
(a) STM image taken from Ref. \cite{kraf97}.
(b) Simulated image representing the surface of constant density with $\rho$ = 5$\times$$10^{-4}$ e/${\rm {\AA}}^3$.
For better comparison with the low-bias STM image, the calculated Fermi level was shifted up by 0.4 eV.
The inset shows a contour map drawn with a uniform increment of 0.05 {\AA}, where filled circles denote the topmost In atoms.
(c) Experimental (Ref. \cite{Kraf97}) and simulated line profiles taken along the arrow marked in (b).
}
\end{figure}


Figure 2 shows the simulated STM image of the HEX2 model, which indeed compares well with the experimental feature of two bright and three weak spots per unit cell \cite{kraf97}. 
In our simulation, two of the three weak spots represent the In-In bonding states rather than individual In atoms, thus accounting for the five (not six) protrusions from six surface In atoms per unit cell.
In Fig. 2(c), the calculated charge corrugation is in quantitative agreement with the STM line profile \cite{Kraf97}: the maximum height difference between the peaks (0.23 {\AA}) and the interval of the two highest peaks (3.32 {\AA}) compare well with the experimental values of 0.25 {\AA} and 3.4 {\AA}, respectively.


It is interesting to compare the topographic heights of the HEX2 model and the $\sqrt{7}$-rect phase.
When simulated at a same bias voltage and charge density, the HEX2 model appears lower by 0.24 {\AA} than the $\sqrt{7}$-rect model, in good agreement with the STM observation that the $\sqrt{7}$-hex phase appears a little lower by 0.1--0.2 {\AA} \cite{kraf95}.
On the other hand, the 1.2 ML single-layer model of Uchida $et$ $al.$ \cite{uchi13} was found to appear far lower by 2.21 {\AA}.


Figure 1(d) displays another fascinating feature of the HEX2 model: it could form an atomically sharp phase boundary with the coexisting $\sqrt{7}$-rect phase.
This boundary matching is achieved by a 60 $^{\circ}$ clockwise rotation of the HEX2 model in Fig. 1(c).
The resulting angular relation matches well the STM observation that, when the $\sqrt{7}$-hex and $\sqrt{7}$-rect phases coexist, the angle difference of their unit cells is 60$^{\circ}$ \cite{kraf95,kraf97}.
Their typical coexistence \cite{kraf95} is also explained by the aforementioned energetics: since the HEX2 and $\sqrt{7}$-rect phases are energetically comparable, both phases would locally develope during the In deposition at high temperatures. 


The HEX2 model, consisting of a quasihexagonal In double layer, is not only energetically stable but also reveals sufficiently sound STM features, so we readily propose it as the long-sought structural model for the $\sqrt{7}$-hex phase and further explore its electronic structure.

\begin{figure}
\centering{ \includegraphics[width=8.0 cm]{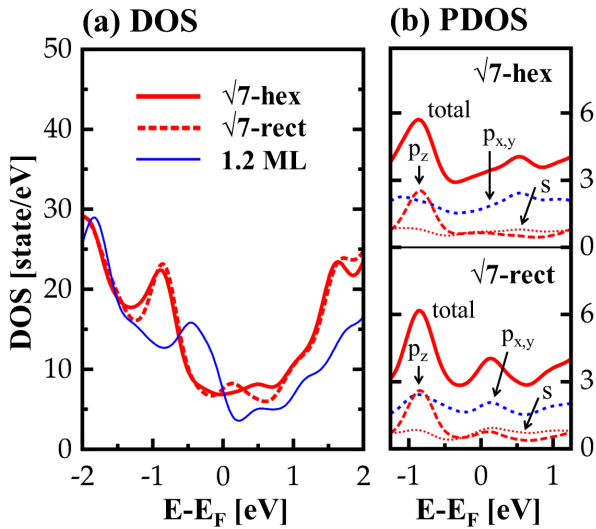} }
\caption{ \label{fig3}
(Color online)
(a) DOS for the present $\sqrt{7}$-hex model, the $\sqrt{7}$-rect model of Ref. \cite{park12}, and the 1.2 ML model of Ref. \cite{uchi13}.
(b) Projected DOS of the In $p$ and $s$ orbitals for $\sqrt{7}$-hex and $\sqrt{7}$-rect. 
}
\end{figure}


Figure 3 shows the calculated density of states (DOS) of our $\sqrt{7}$-hex model.
Noticeable is that the $\sqrt{7}$-hex model reveals almost the same DOS spectrum as the $\sqrt{7}$-rect model, well reflecting their common structural nature (i.e., the same In coverage of 2.4 ML and similar double-layer structures), while the 1.2 ML single-layer model of Uchida $et$ $al.$ \cite{uchi13} does not. 
Our $\sqrt{7}$-hex model is thus compatible with the experimental report that the $\sqrt{7}$-hex and $\sqrt{7}$-rect phases reveal similar scanning tunneling spectra \cite{kraf95}.
There is of course a little difference in the DOS curves, reflecting the underlying structural difference: the small peak at $+$0.28 eV in the $\sqrt{7}$-rect model is a little shifted up to $+$0.53 eV in the $\sqrt{7}$-hex model.  
As seen in Fig. 3(b), the difference stems from the $p_{x}$ and $p_{y}$ orbitals since their lateral interactions are the most affected by the in-plane rectangular-to-hexagonal deformation.  

\begin{figure} 
\centering{ \includegraphics[width=8.0 cm]{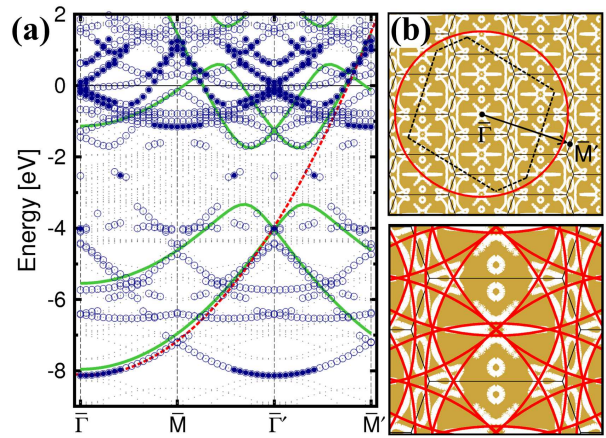} }
\caption{ \label{fig4}
(Color online)
(a) Band structure of $\sqrt{7}$-hex.
Circles denote the In states that contain more than 50\% of charge in the In double layer: the states containing more than 40\% of charge in the top In layer are emphasized by filled circles.
Solid (green) curves denote the band structure of the (1$\times$1) freestanding In double layer.
A dashed (red) curve denotes a parabolic fitting to the bottom part of the In-derived free-electron-like band.
Here, the Fermi level was set to zero.
(b) Fermi contours. 
The ($\sqrt{7}\times\sqrt{3}$) and In-(1$\times$1) Brillouin zones are denoted by solid and dashed lines, respectively.
The arrow denotes the line for the band-structure calculation in (a).
The large (red) circle represents the 2D Fermi circle constructed from the parabolic bands described in (a). 
In the lower panel, the ($\sqrt{7}\times\sqrt{3}$) zone folding of the Fermi circle accounts for most of the complex Fermi contours.
}
\end{figure}


Figure 4 shows the band structure of the $\sqrt{7}$-hex model.
It is metallic with a broad distribution of In-derived states. 
Interestingly, the In band structure could be traced from those of the (1$\times$1) freestanding In double layer [mentioned in Fig. 1(c)]: the ($\sqrt{7}\times\sqrt{3}$) zone folding of this (1$\times$1) result accounts well for the complicated In bands and the band gap between $-$3.36 eV and $-$1.75 eV of the $\sqrt{7}$-hex phase.
A parabolic fitting (with an effective mass of 0.92 $m_{\rm e}$) connects well the In states from the bottom at $-$8.13 eV to over the Fermi level.
These free-electron-like states were found evenly distributed on both of the top and second In layers (not shown here), implying that the double-layer unit may be essential for the 2D free-electron nature.
This surface has similar parabolic bands along other $k$ directions, thereby leading to a circular Fermi surface: Its radii are 1.39 ${\AA}^{-1}$ and 1.36 ${\AA}^{-1}$ along the $k_{\rm x}$ and $k_{\rm y}$ directions, respectively.
Figure 4(b) shows the large Fermi circle in the extended ($\sqrt{7}\times\sqrt{3}$) Brillouin zones, and the actual Fermi contours are mostly accounted for by the the ($\sqrt{7}\times\sqrt{3}$) zone folding of the large Fermi circle, as seen in the bottom panel. 
This 2D free-electron nature of the $\sqrt{7}$-hex phase is similar to that of the $\sqrt{7}$-rect phase: in our calculations, the  $\sqrt{7}$-rect phase has the band bottom of $-$8.18 eV, the effective mass of 0.95 $m_{\rm e}$, and the Fermi circle of $~$1.4 ${\AA}^{-1}$. 


It is worth mentioning that there is one more ($\sqrt{7}\times\sqrt{3}$) hexagonal phase, differing from the present $\sqrt{7}$-hex phase prepared at high temperatures (HT) of about 400 $^{\circ}$C \cite{kraf95, kraf97}.
Saranin $et$ $al.$ \cite{sara06} obtained a hexagonal In/Si(111)-($\sqrt{7}\times\sqrt{3}$) phase by room-temperature (RT) In deposition on In/Si(111)-($\sqrt{3}\times\sqrt{3}$) and identified it with the HT $\sqrt{7}$-hex phase on the basis of similar STM images.
This identification, however, is open to questions.
First, the STM comparison of both phases was not fairly done: the RT phase was examined at high biases of $+$0.5 V and $+$2 V \cite{sara06}, while the HT phase at as low as -0.2 V and -0.12 V \cite{kraf97}.
Indeed, when taken at about $+$2 V, our STM simulations for the $\sqrt{7}$-hex model do not reproduce the STM image of the RT $\sqrt{7}$-hex surface. 
Second, the RT and HT $\sqrt{7}$-hex phases appear differently in topographic height.
The HT phase appears higher by about 2.0 {\AA} in STM topograph than the reference ($4\times1$) phase \cite{kraf95} whereas the RT phase shows only a small height difference of 0.5 {\AA} as found in a recent atomic-force microscopy image \cite{iwat13}, which is a strong implication of different $\sqrt{7}$-hex phases.
Finally, the RT phase has a metastable nature: it transforms into a ($\sqrt{7}\times\sqrt{7}$) phase during cooling in the range from 265 to 225 K \cite{sara06}, while the HT phase is stable at cryogenic temperatures as observed in superconductivity experiments \cite{uchi11, yama13}. 
Thus, as a metastable intermediate phase, the RT phase should be distinguished from the stable HT phase.


In summary, we proposed a double-layer structural model for the hexagonal In/Si(111)-($\sqrt{7}\times\sqrt{3}$) surface, based on the quantitative microscopic and spectroscopic examinations by DFT calculations. 
This double-layer thickness with the In coverage of 2.4 ML is the common structural feature underlying the typical coexistence of the hexagonal and rectangular phases and their similar electronic structures close to a 2D free-electron gas.  
Therefore, the In/Si(111)-($\sqrt{7}\times\sqrt{3}$) surface, either rectangular or hexagonal, does not represent a single-layer limit of metallic In overlayers: such an ultimate 2D limit is yet to achieve.  


This work was supported by the National Research Foundation of Korea (Grant No. 2011-0008907).
The authors acknowledge useful discussions with Tae-Hwan Kim and Han Woong Yeom.

\newcommand{\PR}[3]{Phys.\ Rev.\ {\bf #1}, #2 (#3)}
\newcommand{\PRL}[3]{Phys.\ Rev.\ Lett.\ {\bf #1}, #2 (#3)}
\newcommand{\PRA}[3]{Phys.\ Rev.\ A\ {\bf #1}, #2 (#3)}
\newcommand{\PRB}[3]{Phys.\ Rev.\ B\ {\bf #1}, #2 (#3)}
\newcommand{\RMP}[3]{Rev.\ Mod.\ Phys.\ {\bf #1}, #2 (#3)}
\newcommand{\PST}[3]{Phys.\ Scr.\ T\ {\bf #1}, #2 (#3)}
\newcommand{\PML}[3]{Phil.\ Mag.\ Lett.\ {\bf #1}, #2 (#3)}
\newcommand{\SCI}[3]{Science\ {\bf #1}, #2 (#3)}
\newcommand{\SSA}[3]{Surf.\ Sci.\ {\bf #1}, #2 (#3)}
\newcommand{\SSCO}[3]{Solid\ State\ Comm.\ {\bf #1}, #2 (#3)}
\newcommand{\SSR}[3]{Surf.\ Sci.\ Rep.\ {\bf #1}, #2 (#3)}
\newcommand{\SRL}[3]{Surf.\ Rev.\ Lett.\ {\bf #1}, #2 (#3)}
\newcommand{\NA}[3]{Nature\ {\bf #1}, #2 (#3)}
\newcommand{\NAT}[3]{Nature\ Phys.\ {\bf #1}, #2 (#3)}
\newcommand{\JP}[3]{J.\ Phys.\ {\bf #1}, #2 (#3)}
\newcommand{\JACS}[3]{J.\ Am.\ Chem.\ Soc.\ {\bf #1}, #2 (#3)}
\newcommand{\JAP}[3]{J.\ Appl.\ Phys.\ {\bf #1}, #2 (#3)}
\newcommand{\JCP}[3]{J.\ Chem.\ Phys.\ {\bf #1}, #2 (#3)}
\newcommand{\JPCS}[3]{J.\ Phys.\ Chem.\ Solids.\ {\bf #1}, #2 (#3)}
\newcommand{\JVSA}[3]{J.\ Vac.\ Sci.\ Technol.\ A\ {\bf #1}, #2 (#3)}
\newcommand{\JJAP}[3]{Jpn.\ J.\ Appl.\ Phys.\ {\bf #1}, #2 (#3)}
\newcommand{\ASS}[3]{Appl.\ Surf.\ Sci.\ {\bf #1}, #2 (#3)}
\newcommand{\APL}[3]{Appl.\ Phys.\ Lett.\ {\bf #1}, #2 (#3)}
\newcommand{\APE}[3]{Appl.\ Phys.\ Exp.\ {\bf #1}, #2 (#3)}
\newcommand{\BBPC}[3]{Ber.\ Bunsen-Ges.\ Phys.\ Chem.\ {\bf #1}, #2 (#3)}
\newcommand{\CPL}[3]{Chem.\ Phys.\ Lett.\ {\bf #1}, #2 (#3)}
\newcommand{\LTP}[3]{Low\ Temp.\ Phys.\ {\bf #1}, #2 (#3)}
\newcommand{\TSF}[3]{Thin\ Solid\ Filims\ {\bf #1}, #2 (#3)}
\newcommand{\VAC}[3]{Vacuum\ {\bf #1}, #2 (#3)}
\newcommand{\EL}[3]{Europhys.\ Lett.\ {\bf #1}, #2 (#3)}
\newcommand{\IJMPB}[3]{Int.\ J.\ Mod.\ Phys.\ B.\ {\bf #1}, #2 (#3)}
\newcommand{\CJCP}[3]{Chin.\ J.\ Chem.\ Phys.\ {\bf #1}, #2 (#3)}
\newcommand{\NRL}[3]{Nano.\ Res.\ Lett.\ {\bf #1}, #2 (#3)}
\newcommand{\PLA}[3]{Phys.\ Lett.\ A.\ {\bf #1}, #2 (#3)}

\end{document}